# Mapping and Evolving Interoperability Testing in European Energy Systems: The int:net Perspective

Thomas I. Strasser
*Power & Renewable Gas Systems – Center for Energy*
AIT Austrian Institute of Technology
Vienna, Austria
0000-0002-6415-766X

Edmund Widl
*Power & Renewable Gas Systems – Center for Energy*
AIT Austrian Institute of Technology
Vienna, Austria
0000-0002-2834-306X

Carlos Ayon Mac Gregor
B.A.U.M. Consult GmbH
Berlin, Germany
0009-0003-6295-6040

Mirko Ginocchi
Institute for Automation of Complex Power Systems
RWTH Aachen University
Aachen, Germany
0000-0002-8867-7504

René Kuchenbuch
Energy Division
OFFIS e.V.
Oldenburg, Germany
0009-0000-0872-4722

*Abstract*—The ongoing transformation of the European energy landscape, driven by the integration of renewable energy sources, digital technologies, and decentralized systems, requires a high degree of interoperability across diverse components and systems. Ensuring that these elements can exchange information and operate together reliably is essential for achieving a secure, flexible, and efficient energy supply infrastructure. While several initiatives have contributed to the development of smart grid testing infrastructures, they do not provide a dedicated or comprehensive focus on interoperability testing. A structured and harmonized overview of interoperability testing capabilities across Europe is therefore still missing. This work therefore presents a novel contribution by analyzing the European interoperability testing facility landscape through a structured survey of 30 facilities. It provides a categorized inventory of testing infrastructures, applied methodologies, and reference test cases, and introduces a blueprint for the development of future testing environments. The findings contribute to the establishment of a coordinated European ecosystem for interoperability testing, supporting collaboration, innovation, and alignment with the goals of the energy transition.

## I. Introduction

Interoperability is a critical element for the successful implementation of diverse applications and services within power and energy systems, especially related to the ongoing energy transition [1]–[3]. It necessitates the seamless collaboration of various methodologies, solutions, components, and devices. Achieving interoperability between interconnected systems can be accomplished through several strategies, such as (i) requirements engineering, (ii) interoperability-by-design, (iii) the use of reference architectures, and (iv) compliance with established standards and guidelines [4]–[6].

However, the mere application of these principles is not sufficient to ensure interoperability in smart grids and energy systems. It is mandatory to conduct thorough interoperability tests and issue corresponding testing reports and certificates that validate the successful completion of these tests. The landscape of interoperability testing is characterized by a diversity of procedures, categorizations, evaluations, and assessment criteria, reflecting the varied interests across different initiatives and stakeholders [5], [6].

While several initiatives such as the Joint Research Centre (JRC) of the European Commission (EC) Smart Grid Laboratory Inventory [7], the DERlab[1] and IEA ISGAN/SIRFN[2] networks, and projects like ERIGrid and ERIGrid 2.0[3] have contributed significantly to the development of smart grid testing infrastructures, they do not provide a dedicated or comprehensive focus on interoperability testing. These efforts often emphasize component-level validation or general laboratory capabilities, leaving a gap in system-level interoperability testing and harmonized assessment frameworks. A comprehensive synthesis and repository of interoperability testing methodologies and facilities across Europe is therefore still missing [5], [6].

This work addresses these gaps by presenting a novel and structured analysis of the European interoperability testing facility landscape. It builds on the outcomes of the Horizon Europe project *int:net*[4], which provides a detailed inventory of interoperability testing infrastructures, applied methodologies, and reference test cases and profiles [8]. The findings are based on a targeted survey and categorization process. Additionally, a blueprint for the development of future testing facilities is in-

This work received funding from the European Union's Horizon Europe Research and Innovation Programme under grant agreement N°101070086 (*"Interoperability Network for the Energy Transition" (int:net)*).

[1]https://der-lab.net
[2]https://www.iea-isgan.org/our-work3/wg_5/
[3]https://erigrid2.eu
[4]https://intnet.eu

troduced, outlining key technical, organizational, and business requirements. The results aim to support the advancement of a coordinated and harmonized European interoperability testing ecosystem for energy systems.

The remaining parts of this paper are organized as follows. Section II describes the data collection and analysis approach. Key findings from the inventory are presented in Section III. Section IV outlines a blueprint for future testing facilities. A discussion of implications and opportunities follows in Section V, and Section VI concludes the paper.

## II. METHODOLOGY

To assess the landscape of interoperability testing in European energy systems, a structured four step methodology was applied as depicted in Fig. 1 and described below. The approach aimed to identify relevant testing facilities, analyze their capabilities, and derive an inventory that includes testing approaches and reference test cases.

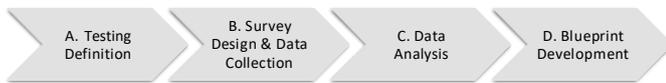

Fig. 1. Overview of the applied methodology.

### A. Definition of Interoperability Testing

Interoperability testing in the context of this work, as defined by the JRC [9], and the SMARTGRIDS Austria IES-Process [10], is a system-level evaluation that verifies whether different systems or products can exchange information and function together as intended. Interoperability testing therefore assesses the ability of two or more systems, components, or devices to exchange information and to make mutual use of the information exchanged. Prior to this, conformance testing ensures that each system, component, or device meets specific standards. While conformance testing confirms individual compliance, interoperability testing focuses on the interaction between systems to ensure they work together effectively.

### B. Survey Design, Structure, and Data Collection

The primary data source for this work was a comprehensive online survey conducted via the European Union (EU) Survey platform[5]. Developed within the scope of *int:net*, the survey aimed to collect detailed information on European research infrastructures, laboratories, and testing facilities engaged in interoperability testing within the energy sector. It was structured into the following six main parts:

- *Part 1: Test Facility Information* – General details such as name, location, year of establishment, legal status, field of expertise, and type of facility (e.g., single-sited, distributed, event-based).
- *Part 2: Testing Focus and Methodology* – Domains and technologies tested, use cases, standards alignment, and applied testing methodologies.
- *Part 3: Challenges, Solutions, and Quality Assurance* – Key challenges, strategies for overcoming them, and quality assurance practices.
- *Part 4: Client Engagement and Feedback* – Types of clients, involvement in testing processes, and feedback mechanisms.
- *Part 5: Future Topics and Interests* – Anticipated trends, interest in the *int:net* label, and the Interoperability Management and Audit System (IntMAS)[6].
- *Part 6: Final Inquiries* – Additional insights, collaboration practices, and support needs.

The questionnaire with 39 questions in total was designed to capture a wide range of information, including fields of expertise, offered services, involvement in interoperability testing, applied testing approaches, reference test cases and profiles, and future interests. The estimated completion time was 20–30 minutes.

Potential participants were selected based on a review of existing inventories (incl. the JRC Smart Grid Laboratory Inventory [7]), domain-specific funded projects such as ERIGrid/ERIGrid 2.0, laboratory networks including DERlab, IEA ISGAN/SIRFN, and others. The survey was fully compliant with European data protection regulations.

### C. Data Analysis and Categorization

The collected responses were analyzed and categorized to form the basis of the inventory. Facilities were grouped according to their interoperability testing capabilities, testing methodologies, and use of reference test cases. The analysis also considered the type of infrastructure (physical, virtual, or hybrid) and the level of system integration addressed (component, subsystem, or system level).

Testing approaches were classified according to the work in [6] into four categories:

- Methodologies,
- Frameworks and tools,
- Standards, guidelines, and policies, and
- Other approaches from the literature.

Reference test cases and profiles were also compiled and linked to the corresponding testing approaches and Smart Grid Architecture Model (SGAM) domains and zones [11], [12].

### D. Blueprint Development

Based on the insights gained from the inventory, a blueprint for the development of new interoperability testing facilities was derived. This blueprint includes technical, organizational, and business recommendations to support the establishment of future-ready testing environments. It is intended to guide stakeholders in designing facilities that align with evolving interoperability requirements in the energy sector.

## III. RESULTS AND ANALYSIS

A total of 30 European testing facilities provided input to the survey regarding their interoperability testing capabilities.

---

[5]https://ec.europa.eu/eusurvey/runner/intnet-iop-test-facilities-survey

[6]https://intnet.eu/images/IntMAS V 0.9.pdf

This corresponds to a return rate of 43.5% based on the organizations contacted and 31.3% based on individual contacts, which is considered an acceptable outcome. Of these facilities, 21 (i.e., 70%) indicated active engagement in interoperability-related topics and testing activities. Furthermore, 20 facilities reported the use of interoperability testing approaches, concepts, and tools. However, only 8 facilities utilize automated testing tools, and just 10 conduct their testing activities in accordance with national or international standards, of which only 3 have formal certification or accreditation processes in place. Additionally, 15 facilities offer customized interoperability testing services, while 4 do not, and 11 did not respond to this question. The survey also revealed varying levels of interest in the IntMAS: 5 facilities expressed clear interest, 12 indicated potential interest, one facility was not interested, and the remaining did not respond.

These insights provided a robust foundation for categorizing the facilities and compiling a structured inventory, organized into the following three main categories:

A. "Testing Facilities",
B. "Testing Approaches & Tools", and
C. "Reference Test Cases & Profiles".

The resulting inventory is publicly available via *int:net's* Zenodo Community[7]. It's main categories are described below.

### A. Overview of Testing Facilities

The inventory reveals a diverse and geographically distributed landscape of testing infrastructures across Europe, covering domains such as smart grids, distributed energy resources (DER), power electronics, energy management systems, and electric mobility. Facilities vary in type, including:

- *Single-sited laboratories* – e.g., AIT the SmartEST Laboratory, Dynamic Power Systems Laboratory (DPSL) at the University of Strathclyde,
- *Distributed infrastructures* – e.g., DNV's multi-location test environments, JRC's interoperability test laboratories in Ispra (Italy) and Petten (Netherlands), and
- *Event-based setups* – plugfests, connectathons, etc.

These 21 facilities support different levels of system integration, from component-level validation to full system-level evaluations. The majority are publicly funded, while others operate under private or public-private partnership (PPP) models as depicted in Fig. 2.

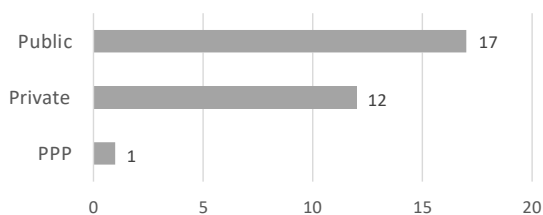

Fig. 2. Legal status of facilities.

Table I summarizes the metadata collected per facility.

[7]https://doi.org/10.5281/zenodo.15084185

TABLE I
STRUCTURE OF THE "TESTING FACILITIES" TABLE.

| Field | Description |
|---|---|
| Name | Name of the testing facility |
| Organization | Responsible institution |
| Location | City and country |
| Year Established | Year of foundation |
| Legal Status | Public, Private, or PPP |
| Field of Expertise | Domains such as DER, customer, etc. |
| Offered Services | Testing, validation, certification, etc. |
| Type of Facility | Single-sited, distributed, event-based |
| Interoperability Testing | Yes/No |
| Used Approaches | Testing approaches used by the facility |
| Reference TCs | Reference test cases used by the facility |

### B. Applied Testing Approaches and Tools

The analysis identified 22 distinct testing approaches and tools, categorized into four groups [6]:

- *Methodologies* – e.g., JRC Smart Grid Interoperability Testing Methodology, ERIGrid Holistic Test Description (HTD), SMARTGRIDS Austria IES Process,
- *Frameworks and Tools* – e.g., AIT Virtual Verification Laboratory (VLab), IHE Gazelle, JRC's SG-DoIT,
- *Standards and Guidelines* – e.g., ENTSO-E CGMES Conformity Assessment Framework, NIST Smart Grid Interoperability Framework, and
- *Other Approaches* – e.g., semantic web practices, IoT-specific testing models.

These approaches vary in maturity and adoption. While some facilities rely on established frameworks, others use custom or domain-specific methods as outlined in Fig. 3.

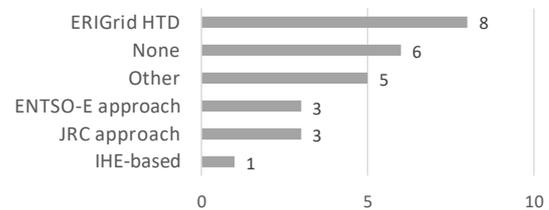

Fig. 3. Legal status of facilities.

### C. Implemented Reference Test Cases and Profiles

A total of 16 reference test cases (TCs) and interoperability profiles were reported. These include:

- ERIGrid 2.0 test cases (e.g., TC16, TC24),
- SMARTGRIDS Austria IES profiles (e.g., TF-LEC, TF-VPP, TF-EV), and
- ENTSO-E CIM/CGMES interoperability test cases.

The test cases span multiple SGAM domains (e.g., DER, distribution, customer) and zones (e.g., field, station, operation). Table II summarizes the structure of the "Reference Test Cases & Profiles" table of the inventory.

TABLE II
STRUCTURE OF THE "REFERENCE TEST CASES & PROFILES" TABLE.

| Field | Description |
| --- | --- |
| Name | Test case or profile name |
| Organization | Responsible entity |
| Start/Update | Development timeline |
| Related Approach | Linked methodology or tool |
| SGAM Domain/Zone | Coverage of SGAM layers |
| Description | Scope and purpose |

*D. Key Observations*

The following key insights were derived from the analysis:

- *Growing relevance:* Interoperability testing is gaining traction, especially in DER and smart grid contexts.
- *System-level gap:* Most testing remains at the component level; system-level validation is still underdeveloped.
- *Methodological diversity:* A wide range of approaches is used, reflecting facility specialization but also a lack of harmonization.
- *Tool adoption:* Only 8 facilities reported using automated testing tools; formal certification is rare.
- *Future orientation:* Many facilities expressed interest in adopting structured frameworks like IntMAS and the *int:net* label.

## IV. BLUEPRINT FOR FUTURE FACILITIES

Based on the analysis of existing testing infrastructures and the identified gaps in current practices, a blueprint for the development of future interoperability testing facilities has been derived. This blueprint outlines key technical, organizational, and business requirements to support the establishment of robust and future-ready testing environments. It builds on the findings of the *int:net* inventory and incorporates recommendations [8].

*A. Technical Requirements*

Future facilities should be designed to support both physical and virtual testing environments. This includes the integration of advanced platforms such as real-time simulation systems, Hardware-in-the-Loop (HIL) setups, and digital twins. Scalability and modularity are essential to accommodate evolving technologies and diverse integration scenarios.

Facilities should enable testing across multiple levels of system integration—component, subsystem, and full system level. To support comprehensive interoperability validation, facilities should also maintain repositories of reference test cases and profiles aligned with SGAM domains and zones:

- Support for hybrid infrastructures (physical and digital),
- Capability for automated and remote testing,
- Integration of standardized testing methodologies (e.g., JRC, IES Process, ERIGrid HTD),
- Use of reference architectures and test case libraries.

*B. Organizational and Administrative Aspects*

Effective operation of interoperability testing facilities requires clear access and usage policies, efficient scheduling systems, and comprehensive user support services. Collaboration with other laboratories, research institutions, and industry stakeholders is essential to foster knowledge exchange and methodological alignment.

Facilities should actively participate in communities such as the *int:net* Interoperability Focus Groups (IFGs) and offer training programs, documentation, and workshops to promote best practices:

- Transparent access and usage policies,
- Stakeholder engagement and cross-facility collaboration,
- Training programs and technical documentation, and
- Participation in interoperability maturity assessments (e.g., *int:net's* Interoperability Maturity Model (IMM) EMINENT [13]).

*C. Business Considerations*

A sustainable business model should combine public funding with service-based revenue streams. Facilities should offer a portfolio of services including interoperability, conformance, and performance testing, but also consultancy and certification.

Participation in labeling schemes such as the *int:net* label and adoption of frameworks like the IntMAS can enhance credibility, visibility, and benchmarking capabilities:

- Mixed funding models (public and private),
- Service offerings tailored to industry and research needs,
- Certification and labeling to ensure quality assurance, and
- Strategic branding and market positioning.

## V. DISCUSSION

The results of this work highlight both the strengths and limitations of the current interoperability testing landscape in Europe. While a number of well-established facilities exist, the diversity of testing approaches and the limited focus on system-level validation indicate a clear need for greater harmonization and strategic coordination.

The inventory confirms that many facilities operate in isolation, applying custom or domain-specific methodologies. This fragmentation limits the comparability of results, the scalability of testing practices, and the potential for cross-border collaboration. The adoption of shared frameworks, such as the ERIGrid HTD, the IES Process, or the JRC Smart Grid Interoperability Testing Methodology, can support methodological alignment and improve interoperability assurance across the sector [5], [6].

Moreover, the limited use of automated testing tools and formal certification processes suggests that there is significant room for improvement in terms of efficiency, repeatability, and quality assurance. Only a minority of facilities currently employ automated tools or have accreditation mechanisms in place, which hinders the broader adoption of standardized testing practices.

The growing interest in structured frameworks like the IntMAS and the *int:net* label reflects a broader demand for transparency, benchmarking, and continuous improvement. These initiatives offer promising pathways to enhance trust, comparability, and visibility of interoperability testing activities across Europe.

The blueprint proposed in this paper provides a practical foundation for addressing these challenges. It supports the development of new facilities that are aligned with current needs and future trends, while also offering guidance for the enhancement of existing infrastructures. By integrating technical, organizational, and business considerations, the blueprint encourages the creation of robust, scalable, and harmonized testing environments that can accelerate the energy transition and foster innovation.

Future work should focus on expanding the inventory, refining the blueprint based on stakeholder feedback, and supporting the implementation of tools such as IntMAS and EMINENT. Strengthening collaboration through platforms like the *int:net* Community[8] and its IFGs will be essential to drive convergence and shared progress in this critical domain.

## VI. Conclusions

This paper presents a structured overview of interoperability testing capabilities in European energy systems, based on the outcomes of *int:net*. Through a detailed survey and analysis, an inventory of testing facilities, applied methodologies, and reference test cases has been compiled and categorized.

The findings reveal a diverse and evolving landscape of testing infrastructures, with significant variation in technical focus, methodological maturity, and service offerings. While many facilities demonstrate strong capabilities at the component level, system-level interoperability testing remains underdeveloped. The inventory also highlights a lack of harmonization in testing approaches, limited use of automation, and a general absence of formal certification processes.

To address these challenges and support the development of a coordinated European testing ecosystem, a blueprint for future interoperability testing facilities has been introduced. This blueprint outlines key technical, organizational, and business requirements to guide the establishment and enhancement of testing environments that are robust, scalable, and aligned with emerging needs.

The results contribute to ongoing efforts to foster collaboration, transparency, and methodological alignment across the European energy sector. Future work will focus on:

- Implementing the proposed blueprint in new and existing facilities,
- Further developing supporting tools such as the IntMAS,
- Expanding and refining the inventory of testing facilities and approaches,
- Promoting the adoption of shared frameworks and reference test cases,
- Strengthening community engagement through platforms like the *int:net* IFGs.

---

[8]https://community.intnet.eu

These efforts aim to accelerate the transition toward a harmonized and high-quality interoperability testing ecosystem that supports innovation, standardization, and the broader goals of the European energy transition.

## Acknowledgment

The authors thank all survey participants for their valuable input, which was essential to developing the inventory and insights presented in this paper.

During the preparation of this work, the authors used Microsoft Copilot to assist with language and grammar corrections as well as stylistic refinement. After using this tool, the authors carefully reviewed and edited the content as needed and take full responsibility for the content of the publication.